
\def\page{\footline={\ifnum\pageno=1 \hfil
          \else\hss\tenrm\folio\hss\fi}}

\def\ahead#1\smallskip{{\vfil\eject{\centerline{\bf#1}}
\smallskip}{\message#1}}
\def\bhead#1\smallskip{{\bigbreak{\centerline{\it#1}}\smallskip}
{\message#1}}
\def\chead#1\par{{\bigbreak\noindent{\it#1}\par} {\message#1}}

\def\head#1. #2\par{\medbreak\centerline{{\bf#1.\enspace}{\it#2}}
\par\medbreak}

\def\levelone#1~ ~ #2\smallskip{\noindent#1~ ~ {\bf#2}\smallskip}
\def\leveltwo#1~ ~ #2\smallskip{\noindent#1~ ~ {\it#2}\smallskip}
\def\levelthree#1~ ~ #2\smallskip{\noindent#1~ ~ {#2}\smallskip}

\def\m{^m\kern-7pt .\kern+3.5pt}
\def\p{^{\prime\prime}\kern-2.1mm .\kern+.6mm}
\def\pone{^{\prime}\kern-1.05mm .\kern+.3mm}
\def\dpoint{^d\kern-1.05mm .\kern+.3mm}
\def\hpoint{^h\kern-2.1mm .\kern+.6mm}
\def\y{^y\kern-1.05mm .\kern+.3mm}
\def\s{^s\kern-1.2mm .\kern+.3mm}

\def\apgt{\ {\raise-.5ex\hbox{$\buildrel>\over\sim$}}\ }
\def\aplt{\ {\raise-.5ex\hbox{$\buildrel<\over\sim$}}\ }
\def\deg{^{\circ}}

\def\hup{^{h}\kern-2.1mm .\kern+.6mm}

%
%

%
%

%
%
\def\today{\number\day\space\ifcase\month\or
January\or February\or March\or April\or May\or June\or July\or
August\or September\or October\or November\or December\fi
\space\number\year}

\def\sqr#1#2{{\vcenter{\hrule height.#2pt
\hbox{\vrule width.#2pt height#1pt \kern#1pt
\vrule width.#2pt}
\hrule height.#2pt}}}

\newcount\equationnumber
\newbox\eqprefix
\def\neweqprefix#1{\global\equationnumber=1
\global\setbox\eqprefix=\hbox{#1}}

\def\autono{(\copy\eqprefix\number\equationnumber)
\global\advance\equationnumber by 1}

\def\trule{\vskip6pt\hrule\vskip2pt\hrule\vskip6pt}
\def\mrule{\noalign{\vskip6pt\hrule\vskip6pt}}
\def\brule{\noalign{\vskip6pt\hrule}}

%
%
%
%


\newcount\refno
\refno=0

\def\beginorefs\par{\begingroup\parindent=12pt
\frenchspacing \parskip=1pt plus 1pt minus 1pt
\interlinepenalty=1000 \tolerance=400 \hyphenpenalty=10000
\everypar={\item{\the\refno.}\hangindent=2.6pc}

\def\nature##1,{{\it Nature}, {\bf##1},}

\def\aa##1,{{\it Astr.\ Ap.,\ }{\bf##1},}
\def\aapr{{\it Astr.\ Ap.,\ }in press.}
\def\ajaa##1,{{\it Astron.\ Astrophys.,\ }{\bf##1},}
\def\ajaapr{{\it Astron.\ Astrophys.,\ }in press.}

\def\aalet##1,{{\it Astr.\ Ap.\ (Letters),\ }{\bf##1},}
\def\aaletpr{{\it Astr.\ Ap.\ (Letters),\ }in press.}
\def\ajaalet##1,{{\it Astron. Astrophys. (Letters),\ }{\bf##1},}
\def\ajaaletpr{{\it Astron. Astrophys. (Letters),} in press.}

\def\aasup##1,{{\it Astr. Ap. Suppl.,\ }{\bf##1},}
\def\aasuppr{{\it Astr.\ Ap.\ Suppl.,\ }in press.}
\def\ajaasup##1,{{\it Astron. Astrophys. Suppl.,\ }{\bf##1},}
\def\ajaasuppr{{\it Astron.\ Astrophys.\ Suppl.,\ }in press.}

\def\aass##1,{{\it Astr. Ap. Suppl. Ser.,\ }{\bf##1},}
\def\aasspr{{\it Astr. Ap. Suppl. Ser.,} in press.}

\def\aj##1,{{\it A.~J.,\ }{\bf##1},}
\def\ajpr{{\it A.~J.,\ }in press.}
\def\ajaj##1,{{\it Astron.~J.,\ }{\bf##1},}
\def\ajajpr{{\it Astron.~J.,} in press.}

\def\apj##1,{{\it Ap.~J.,\  }{\bf##1},}
\def\apjpr{{\it Ap.~J.,} in press.}
\def\ajapj##1,{{\it Astrophys.~J.,\ }{\bf##1},}
\def\ajapjpr{{\it Astrophys.~J.,} in press.}

\def\apjlet##1,{{\it Ap.~J. (Letters),\ }{\bf##1},}
\def\apjletpr{{\it Ap.~J. (Letters),} in press.}
\def\ajapjlet##1,{{\it Astrophys. J. Lett.,\ }{\bf##1},}
\def\ajapjletpr{{\it Astrophys. J. Lett.,} in press.}

\def\apjsup##1,{{\it Ap.~J.~Suppl.,\ }{\bf##1},}
\def\apjsuppr{{\it Ap.~J.\ Suppl.,} in press.}
\def\ajapjsup##1,{{\it Astrophys. J. Suppl.,\ }{\bf##1},}
\def\ajapjsuppr{{\it Astrophys. J.\ Suppl.,} in press.}

\def\araa##1,{{\it Ann. Rev. A.~A.,\ }{\bf##1},}
\def\araapr {{\it Ann. Rev. A.~A.,} in press.}

\def\baas##1,{{\it B.A.A.S.,\ }{\bf##1},}
\def\baaspr{{\it B.A.A.S.,} in press.}

\def\mnras##1,{{\it M.N.R.A.S.,\ }{\bf##1},}
\def\mnraspr{{\it M.N.R.A.S.,} in press.}
\def\ajmnras##1,{{\it Mon. Not. R. Astron. Soc.,\ }{\bf##1},}
\def\ajmnraspr{{\it Mon. Not. R. Astron. Soc.,} in press.}

\def\pasp##1,{{\it Pub.~A.S.P.,\ }{\bf##1},}
\def\pasppr{{\it Pub.~A.S.P.,} in press.}
\def\ajpasp##1,{{\it Publ. Astron. Soc. Pac.,\ }{\bf##1},}
\def\ajpasppr{{\it Publ. Astron. Soc. Pac.,} in press.}
}

\def\tenpoint{
  \font\teni=cmmi10
  \font\ten=cmsy10
  \font\teni=cmmi10
  \font\sevensy=cmsy7
  \font\fivei=cmmi5
  \font\fivesy=cmsy5
  \font\it=cmti10
  \font\bf=cmb10
  \font\bi=cmbi10
  \font\sl=cmsl10
  \textfont0= \tenrm \scriptfont0=\sevenrm
\scriptscriptfont0=\fiverm
  \def\rm{\fam0 \tenrm}
  \textfont1=\teni  \scriptfont1=\seveni
\scriptscriptfont1=\fivei
  \def\mit{\fam1 } \def\oldstyle{\fam1 \teni}
  \textfont2=\tensy \scriptfont2=\sevensy
\scriptscriptfont2=\fivesy
\def\doublespace{\baselineskip=24pt\lineskip=0pt
\lineskiplimit=-5pt}
\def\singlespace{\baselineskip=12pt\lineskip=0pt
\lineskiplimit=-5pt}
\def\oneandahalf{\baselineskip=18pt\lineskip=0pt
\lineskiplimit=-5pt}
}

\def\eighteenpoint{
  \font\eighteeni=cmmi10 scaled\magstep3
  \font\eighteensy=cmsy10 scaled\magstep3
  \font\eighteenrm=cmr10 scaled\magstep3
  \font\twelvei=cmmi12
  \font\twelvesy=cmsy12
  \font\teni=cmmi10
  \font\tensy=cmsy10
  \font\seveni=cmmi7
  \font\sevensy=cmsy7
  \font\it=cmti10 scaled \magstep3
  \font\bf=cmb10 scaled \magstep3
  \font\sl=cmsl10 scaled \magstep3
  \textfont0= \eighteenrm \scriptfont0=\twelverm
\scriptscriptfont0=\tenrm
  \def\rm{\fam0 \eighteenrm}
  \textfont1=\eighteeni  \scriptfont1=\twelvei
\scriptscriptfont1=\teni
  \def\mit{\fam1 } \def\oldstyle{\fam1 \eighteeni}
  \textfont2=\eighteensy \scriptfont2=\tensy
\scriptscriptfont2=\sevensy
\def\doublespace{\baselineskip=30pt\lineskip=0pt
\lineskiplimit=-5pt}
\def\singlespace{\baselineskip=20pt\lineskip=0pt
\lineskiplimit=-5pt}
\def\oneandahalf{\baselineskip=25pt\lineskip=0pt
\lineskiplimit=-5pt}
\def\deg{^{\raise2pt\hbox{$\circ$}}}}


\baselineskip=15pt
\magnification=1200
\hsize=15.0truecm
\vsize=23truecm
\overfullrule=0pt
SUMMARY
\medskip\noindent
CCD photometry in the V band is presented for 7 field RR Lyrae stars selected
from a sample
of eight variables which, according to
data collected in the literature, are expected to be {\it ab}-type
pulsators, to have short periods and hence high metallicity,
and to be located at high {\it z} from
the galactic plane. New periods and epochs are derived
for them.
The new periods are only slightly shorter than the values published on the
last edition of the General Catalog of Variable Stars (GCVS4).
Instead, in six cases our amplitude of
the light variation is significantly smaller than that published on
the GCVS4, and in at least three cases the actual pulsation appears to be
in the first harmonic rather than in the fundamental mode. All the suggested
{\it c}-type pulsators show variations in the amplitude and/or quite
scattered light curves. Possible explanations are given.
\par\noindent
{}From a spectro-photometric analysis of the sample, only DL Com is confirmed
to pulsate in the fundamental mode, to have short period, and to be
located at relatively high {\it z}. A single object cannot be taken
as evidence for a significant metal rich
population at large distance from the galactic plane.
\bigskip
\vfill\eject
\noindent
1. INTRODUCTION
\medskip\noindent
RR Lyrae stars have early been recognized as very good tools to study the
structure
and the evolution of the Galaxy. Their pulsational properties, combined with
their almost constant absolute magnitude, allow one to use them
as tracers of the
different stellar populations evolving in different galactic regions. It is
well known, for instance, that the period distributions of field RR Lyrae
variables
depend on their galactic location (Preston 1959, Layden 1993).
In particular, halo RR Lyraes of Bailey type {\it ab}
show a peaked period-frequency distribution which is a mixture
of those derived from variables located in Oostheroff I and II globular
clusters, bulge RR Lyraes show
the typical asymmetric histogram of Oosterhoff I clusters with no fundamental
pulsators with period shorter than P=0.45$^d$, and disc RR Lyraes show instead
a quite disperse distribution with many short period pulsators in the
fundamental mode (Castellani et al. 1981).
These different behaviours can be
interpreted in terms of different stellar initial metallicity, since several
authors (e.g. Lub 1977) have suggested an anticorrelation between
period and metallicity in {\it ab}-type RR Lyraes. Such anticorrelation has
been
questioned by some other authors (Layden 1993 and references therein) because
of the difficulty in reconciling a period-metallicity relation with the
observed distribution of the parent non variable stars. Despite the large
dispersion of the observational data, however, the dependence of the period
on [Fe/H] is apparent also in their own data (see e.g. Layden's fig.6.4).
\par
{}From a sample of 273 field RR Lyrae stars of Bailey type {\it ab} selected
from the
tape version of the third supplement to the third
edition of the General Catalogue of Variable Stars (Kukarkin et al. 1976,
hereinafter GCVS$_{t}$) with the requirement of having galactic latitude
$\mid$b$\mid\geq\pm45^o$, Castellani et al. (1983, hereinafter CMT) found that
the minimum periods of RR{\it ab} increases with increasing height $z$ on the
galactic plane following the relation
$\Delta log P_{min}/\Delta |z| = 0.004$ dex kpc$^{-1}$. Assuming Lub's (1977)
anticorrelation between period and metallicity, they suggested that the period
increase with height on the galactic plane corresponds to a metallicity
gradient $\Delta [Fe/H]/\Delta |z| = -0.02$ dex kpc$^{-1}$. Both period and
metallicity gradients with $z$ are rather uncertain due to the large errors
affecting the data, but there is no doubt (see also Layden 1993) that short
period and metal rich variables are concentrated within less than 1 kpc from
the galactic plane, whereas long period and metal poor variables have no
apparent concentration and reach quite large $z$.
\par
However, CMT found a few variables deviating from the overall $z$-period
distribution, since they showed periods quite shorter than all the others at
$z$ up to almost 10 kpc. If confirmed, short periods at large heights on the
plane would imply that metal rich stars can reach large distances from the thin
disc and that the thick disc is several kpc wide. On the other hand,
apart from RS Boo for which accurate B,V,K light curves have
been published by  Jones et al. (1988), these {\it anomalous} pulsators
have generally very poor photographic light curves (see for instance
Kurochkin 1961, Meinunger
\& Wenzel 1968) and therefore their periods and
magnitudes (i.e. distances) are quite uncertain.
\par
In order to examine in more detail these deviating stars (namely: RS Bootis,
AT Comae, BE Comae, BS Comae, CU Comae, CY Comae, DL Comae and CM Leonis),
and the related question of metal rich objects at large distances from the
galactic plane, we have observed them for several years, deriving accurate
V light curves and periods from the photometric data. To obtain
also direct estimates
of their metallicities we have taken some spectra at minimum light and applied
the method described by Clementini et al. (1991) to infer [Fe/H] from the
equivalent width of the calcium K-line.
\par
The data acquisition and reduction are described in the next section. The
derived features of each variable are described in Section 3  and the
overall conclusions are discussed and summarized in Section 4.
\bigskip\noindent
2. THE DATA
\medskip\noindent
\bigskip\noindent
2.1 Data Acquisition
\medskip\noindent
All the observations have been carried out at the 1.52 m telescope in Loiano
operated by the Bologna Observatory and equipped with an RCA CCD with
$4.3' \times 2.7'$ field of view and a 0.5 arcsec pixel scale,
or with the BFOSC system.
\par
BFOSC (Bologna Faint Object Spectrograph \& Camera) is a focal reducer-type
spectrograph/camera similar in concept to ESO's EFOSC 1 and 2.
The instrument offers the possibility to
switch from direct imaging to spectroscopy in less than one minute.
The detector presently installed on BFOSC is a
Thomson 1k$\times$1k CCD
giving a field of view of $9.6' \times 9.6'$. The pixel scale is 0.56 arcsec.
\par
Finding charts for the variables observed with CCDs are shown in Fig.1,
where the variable
stars are marked by dashes, and the comparison stars by their identification
names (see Table 2). Each chart shows a region of the sky of
$10' \times 10'$.
Since the purpose of our photometric observations was to
obtain accurate light curves
to improve periods, epochs and mean apparent visual magnitudes (and hence
distances) of the program stars, in general we only took
V exposures of our targets.
The variables are all located toward the North Galactic Pole, therefore they
are
observable from late January through early June. Photometric data have been
acquired in the Johnson's V band during 48 nights from early 1989 through mid
1994, with no observations in 1991 for a telescope breakdown.
The number of CCD images taken for each star ranges between a minimum of 68 for
BE Com and a maximum of 177 for CY Com.
A few R and I exposures were also obtained for one of our objects, namely
CU Com.
In Table 1 we summarize
coordinates (Columns 2 and 3), number of
observations (Column 4), and observed time interval (Column 5), for each
variable.
\par
UBVRI observations of RS Boo were obtained at
the same telescope but with a photoelectric photometer and will be analysed
elsewhere (Clementini et al. 1995).
\par
To guarantee a reliable differential measure of the
magnitude variability of our RR Lyrae stars, each CCD field includes at
least a couple of other (hopefully stable) stars with magnitude comparable to
that of the variable.
Coordinates and derived V magnitudes of the primary comparison stars are
listed in Table 2. Column 2 gives the number of the star according
to the
Hubble Space Telescope Guide Star Catalogue (GSC). The first four
digits identify the GSC region which contains the star and the last four
digits identify the star within the region.
\par
To calibrate our photometry 18 primary standard stars taken from Landolt (1982,
1993)
and covering a wide range of B$-$V colours have been observed with the
Johnson's
V filter at various air masses and in several photometric nights. The
photoelectric V magnitudes and B$-$V colours of the observed calibrators
are listed in Table 3.
\par
Spectra have been taken for our candidate RR{\it ab} variables (except RS
Bootis
which was already available) in the spring 1994 when the light curves, and
therefore the epochs of minimum light, were already well defined.
As a further test, direct images of the objects were taken just before
and after the spectrum
exposure to check the exact location of the spectral observation in the
light curve.
However, only
BS Com turned out to be bright enough to allow a sufficient measure of the
equivalent width
of the calcium K-line with the 1.52 m telescope and the
presently available instrumental configuration, and with exposure
times short enough ($\leq$ 45 min) to
remain in the range of minimum light phases.
\bigskip\noindent
2.2 Data Reduction
\medskip\noindent
All the data have been reduced in the IRAF$^1$ environment, including
\footnote {}
{$^1$IRAF is distributed by National Optical Astronomy Observatories, operated
by the Association of Universities for Research in Astronomy Inc. under
contract
with the National Science Foundation.}
debiassing and flat fielding.
On each CCD frame, we have measured the
instrumental magnitude of the variable stars and of at least two comparison
stars in order to derive the light curve of each variable in terms of
differential magnitudes, and sensibly reduce the uncertainty on our results
due to the atmospheric and local conditions. Only the field around AT Com
does not contain more than one reliable comparison star.
Thanks to the isolated location of our targets, the instrumental magnitudes
of all our images have been derived by direct photon counting with aperture
photometry, taking into account the image profiles by means of APPHOT.
Images on the same CCD frame have all been treated
with the same aperture radius to guarantee the consistency of the magnitude
derived for the variable star with those of the comparison stars.
\par
The standard calibrating stars have been measured exactly in the same way.
Extinction coefficients for the photometric nights were derived from
observations at various air masses of both the standard stars and the
comparison
stars of the program variables. The derived V atmospheric extinction
ranges between 0.29 and 0.47 mag, in good agreement with other
photoelectric determinations of the
seasonal mean extinction of the site.
The conversion of our
instrumental magnitudes v to the Johnson standard system is given by:
\par\noindent
V = 0.995v $-$ 7.772 (a) and V = 0.971v $-$ 4.956 (b),
\par\noindent
for the RCA and BFOSC instrumental magnitudes respectively.
\noindent
Calibration (a) is shown in Fig.2.
\par
Since our photometry has been obtained with two different CCDs and
instrumental set-ups, we
have carefully checked the consistency of the calibrations resulting for
the same stars with the two systems.
No appreciable difference has been found
between the two calibrations and between different calibration nights.
In fact, except for AT Com,
the magnitudes attributed to the various comparison stars by the different
calibrating systems and/or in different observing nights,
agree with each other within $\pm$ 0.03 mag.
This can be considered the uncertainty in the zero point of our photometry.
\par
As a further check of the quality of our photometry and to verify the constancy
of the comparison stars, their magnitudes have been
monitored through all the years of observations and turned out to be stable
within $\pm 0.03$ mag.
There is some suspect that the
comparison star of AT Com might be an intrinsic variable, and therefore
its magnitude has been marked by a colon in Table 2.
Unfortunately, we cannot verify our
hypothesis, since no other bright enough comparison star falls in
the CCD field of AT Com.
\bigskip\noindent
3. RESULTS
\medskip\noindent
The light curves resulting from our observations contain an average of
100 points each, covering a significant length of time (up to five
years) and are shown in the top panels of Fig.3. In the corresponding bottom
panels
we have plotted
the residuals around
the mean magnitude
difference between the two brighter comparison stars of each variable.
Their fair constancy shows that the chosen comparison stars
are indeed non variable.
No residuals are shown for AT Com
since only one comparison star falls
within our observational field of view.
The {\it rms} scatter in this difference is
$\pm$0.02 mag for all stars except
the comparison stars of DL Com, which show larger residuals
up to 0.06 mag, because they are quite fainter than all the other objects
and their photometry is undoubtedly poorer. It is interesting to note that in
general the spread in the light curve is smaller than that in the corresponding
residuals of the bottom panel. This is due to the fact that the variable is
always the brightest star in the CCD frame (except in the case of BE Com)
and its magnitude is
thus measured with higher precision. Particularly striking is the case of
DL Com whose light curve is very tight, despite the fact that the
residuals of its comparison stars are the worst of the whole sample.
\par
New periods for the program stars were determined using only our photometric
data, by means of a period search program facility available at the Bologna
Observatory,  which performs a least-square estimate of frequency
following the prescriptions given by Bloomfield (1976).
The new periods are listed in Column 4 of Table 4. Their accuracy is
$\pm$ one unit in the last digit of the listed values.
Also shown in Table 4 are the new epochs of maximum light (Column 3),
the amplitude of the light
variation (Column 5), and the calibrated V magnitudes at maximum light
(Column 6), derived
from the present photometry.
 In the cases of uncertain maximum and/or minimum
(BE Com, BS Com, CU Com, CM Leo), the quoted amplitudes correspond to the
largest observed variation. Values of
the corresponding quantities given by the
GCVS4 are listed in Columns 7,8, and 9, for ease of comparison.
To this purpose we note that the GCVS4 amplitude and magnitude at
maximum light
of CM Leo, CU, CY, DL, AT and BE Com are from light curves in the B filter,
while for BS Com and RS Boo they are from photographic and V photoelectric
data, respectively. Table 4 shows that the new periods are generally in
agreement with the values published on the GCVS4 (where some of the periods
reported in the previous editions of the Catalogue have been corrected).
Our values
are usually shorter, which is not surprising since some of these objects may
have changing periods (see also remarks on the GCVS4). However,
the differences do not exceed 0.00025 days, with the
exception of CU Com and BS Com for which our periods are about 0.010 and
0.005 days shorter than the GCVS4 ones.
\par
A significant discrepancy is found, instead, in the derived amplitudes of
all the objects except CU Com and RS Boo.
In fact, even allowing for the larger amplitude of the B light
curve of an RR Lyrae with respect to its V light curve,
our amplitudes are significantly smaller than those published on the GCVS4.
In some cases even the pulsation mode must be changed from
fundamental to first harmonic. In fact, while all the stars in
Table 4 are classified as Bailey-type {\it ab} by the GCVS4, we have found
that 3 of them look like {\it c}-type pulsators, and that even for BE Com,
whose photometric data are more noisy, shape and amplitude of the light
curve indicate that the star is more likely to pulsate in the first
harmonic than in the fundamental mode.
Most of the photographic light curves from which
amplitudes and Bailey-types published on the GCVS4  were derived have
rather poor
photometric accuracy. In particular, the photographic light curves of
Kurochkin (1961) and Meinunger \& Wenzel (1968)
are very scattered, which makes it extremely difficult to distinguish between
fundamental and first harmonic pulsation mode, while the classification
based on the amplitude of the light curve could also be misleading
if zero point errors exist in the calibration
of the plates. This effect could explain the very large discrepancy between
our amplitudes and those
by Meinunger \& Wenzel (1968) in particular for CM Leo and AT Com (see Sections
3.1, 3.5). We will discuss these issues
more in detail in Section 4.
\par
The photometric data of the program stars are
presented in Table 5 and are also available from the first author
by electronic mail. For each observation we list Heliocentric Julian Day
(HJD; Column 1), phase at mid exposure (Column 2), and calibrated
V magnitudes
(Column 3). Phases in Table 5 have been calculated
according to the ephemerides given in Table 4.
\medskip\noindent
3.1 CM Leonis
\medskip\noindent
The light curve is shown in Fig.3a and contains 71 points observed
in 9 different nights from
February 19 1994 through May 21 1994. The best resulting period is
0.361479$^d$ and the amplitude is 0.49 mag. Both the small amplitude
and the symmetric shape of the light curve suggest that CM Leo is a {\it
c}-type
RR Lyrae, but the curve looks as resulting from the superposition of two well
separated subcurves with $\sim$ 0.1 mag amplitude difference.
The m$_{pg}$ light curve of this star was published by Meinunger \& Wenzel
(1968): their data are very scattered and the amplitude is 1.1 mag instead of
our 0.49 mag. \medskip\noindent
3.2 CU Comae
\medskip\noindent
The light curve is shown in Fig.3b and contains 118 points observed
in 7 different nights from
January 31 1989 through April 29 1994. The best resulting period is
0.405749$^d$ and the maximum amplitude is 0.58 mag, thus indicating that
CU Com is probably a {\it c}-type variable. The region around maximum light
appears splitted in two different branches with amplitude difference of
about 0.15 mag.
The m$_{pg}$ light curve of Meinunger \& Wenzel
(1968) for CU Com is very poor, and has an amplitude of 0.5 mag
which makes it difficult to understand why this star was classified as
{\it ab}-type. We have a few I and R frames (9 and 6, respectively) of
this object covering the phase region : 0.47 $\div$ 0.84, i.e.
from just before minimum light to about 2/3
of the rising branch of the light curve. The corresponding
uncalibrated V$-$R
and V$-$I colour curves have amplitudes and shapes
compatible with CU Com being a {\it c}-type RR Lyrae.
\medskip\noindent
3.3 CY Comae
\medskip\noindent
The light curve is shown in Fig.3c and contains 177 points observed in 16
different nights from March 13 1989 through April 28 1994. The best resulting
period is 0.757880$^d$, in perfect agreement with the GCVS4 value,
and the amplitude is 0.70 mag.
It must be noted that this star was originally included in our sample
since the GCVS$_{t}$ gave for it a much shorter period, P=0.4311$^d$.
Despite the faintness of CY Com, the light
curve is tight and well defined, showing the typical asymmetry of
an {\it ab}-type RR Lyrae.
CCD photometry in the V and R bands has recently been obtained for CY Com
by Schmidt (1991) and Schmidt \& Reiswig (1993). Their light
curves contain 27 data points, shown in Figure 2 of
Schmidt \& Reiswig (1993). Their derived amplitude of the visual light
variation
is A$_{V}$=0.71 mag and the intensity mean magnitude is
$< V >$=14.61. These values compare extremely well with our
values of 0.70 and 14.62 mag, respectively. The V magnitude at minimum light
read from Figure 2 of Schmidt \& Reiswig (1993) is V$_{min} \sim 14.92 -
14.93$ mag and is in good agreement with our value of 14.95 mag.
Schmidt's (1991) time of maximum light for
CY Com is instead rather approximate, mainly because that phase
is not sufficiently well covered in his observations.

\medskip\noindent
3.4 DL Comae
\medskip\noindent
The light curve is shown in Fig.3d and contains 94 points observed in 14
different nights from February 19 1994 through June 18 1994. The best
resulting period, 0.4320590$^d$, is slightly shorter than that given
by the GCVS4 where, however, it is specifically remarked as possibly
varying.
The light curve has an amplitude of 1.26 mag and the typical
asymmetric shape of an {\it ab}-type RR Lyrae.
The photographic light curve of DL Com was published by Kurochkin
\& Pochinok (1974) and has an amplitude of $\sim$ 1.5 mag. Taking
into account the scatter and the different photometric band of their data,
the two amplitudes are consistent with each other, thus
suggesting that the amplitude of 2.0 mag quoted on the GCVS4 is wrong.
\par\noindent
The agreement found for this star, as well as between our data and
those of Smith \& Reiswig (1993) for CY Com,
points out that there may be problems in
the data of Meinunger \& Wenzel (1968) also for the other stars.
\medskip\noindent
3.5 AT Comae
\medskip\noindent
The light curve is shown in Fig.3e and contains 105 points observed in 7
different nights from
February 22 1992 through April 30 1994. The best resulting period is
0.3444676$^d$ and the amplitude is 0.62  mag. The latter value, combined
with the shape of the light curve, suggests that AT Com is more probably
pulsating in the first harmonic than in the fundamental mode. The scatter
at minimum light suggests that the star may suffer from Blazhko effect.
However, we should also recall that we have no check on the stability of our
photometry of this variable, due to the presence of only one measurable
comparison star in its CCD field.
Therefore, we cannot completely exclude that the dispersion
in the region of minimum light might be due either to photometric spread
or to an intrinsic variation of the
comparison star.
The photographic light curve of this star was published by
Meinunger \& Wenzel (1968). While the shapes of the two curves are in good
agreement, the amplitude derived from these previous data is about 1.0 mag
larger than ours.
\medskip\noindent
3.6 BE Comae
\medskip\noindent
The light curve is shown in Fig.3f and contains 68 points observed from
May 5 1989 through June 5 1994. The best resulting period is 0.414209$^d$.
The curve is very symmetric and, since the amplitude is 0.58 mag,
BE Com is most probably a {\it c}-type RR Lyrae. The spread around all the
light curve may be due to the faintness of the variable. However,
the small residuals between the magnitudes of the two comparison stars, which
are only $\sim$0.5 mag brighter, suggests that at least part of the spread is
real, although we do not have a clear explanation for it. Given the
scatter of our light curve around maximum light, we have not attempted
to derive a new epoch from our data and have adopted the value
published on the GCVS$_t$ (i.e. 37705.608). We suspect that a typing error
affects the value of the epoch published on the GCVS4 (i.e. 37705.68).
In fact, if the latter is used the maximum light of
BE Com is shifted backward to phase 0.75. The photographic light curve
of BE Com has been published by Meinunger \& Wenzel (1968). Given the
very large scatter of their data, both shape and amplitude of their
light curve are compatible with BE Com being a {\it c}-type pulsator.
\medskip\noindent
3.7 BS Comae
\medskip\noindent
The light curve is shown in Fig.3g and contains 97 points observed in 13
different nights from April 12 1990 through June 18 1994. The best resulting
period is 0.36296680$^d$ and the amplitude is 0.7 mag.
Smith (1990), on the basis of the
hydrogen spectral type inferred from spectra of BS Com
used in a $\Delta$S
analysis of the star, has reached the conclusion that BS Com is probably a
{\it c}-type pulsator; however our light curve for the star is not
symmetric.
Moreover, the curve appears splitted in two branches
with rather different magnitudes around the phase of maximum light. This
split has no counterpart in the distribution of the magnitude residuals of
the comparison stars and therefore we consider it real. Indeed,
a note on the GCVS4 reports the shape of the light curve of BS Com to vary,
and
Kurochkin (1961) noticed that the strong variation in the light curve
may be explained as due to Blazhko effect.
Kurochkin light curve is very scattered, but its amplitude is in reasonably
good agreement with ours.
A detailed analysis of our data seems to suggest that the star
is an {\it ab}-type variable affected by Blazhko with a periodicity of about
27-28 days.
\par
We have acquired one spectrum of BS Com at phase 0.72.
Direct images of BS Com taken just before
and after the spectrum exposure confirmed that
the spectral observation was exactly located
at minimum
light.
The spectrum has resolution of about 1.4 \AA~ and S/N $\sim$ 20,
and allowed us
to derive
a reliable measure of the calcium K-line equivalent width and to apply the
method described by Clementini et al. (1991) and recently recalibrated
by Clementini et al. (1994). In this way, we derive a very
low metal abundance [Fe/H]$\simeq -$2.0.
Smith's (1990) spectra of BS Com were taken at phases that, according to
our ephemerides, correspond to 0.90 and 0.63 respectively,
and he derived from the latter, which is already at minimum light,
a $\Delta$S value of 8.2.
This value also indicates that the variable is metal poor, and corresponds
to [Fe/H]=$-$1.67 if the metallicity scale by Clementini et al. (1994) is
adopted.
\medskip\noindent
3.8 RS Bootis
\medskip\noindent
This object is the only well studied one of our sample and will be described
by Clementini et al. (1995) in a different context and with additional data.
For our purposes, here it suffices to summarize the resulting features.
RS Boo is an {\it ab}-type RR Lyrae (Spinrad 1959), with short period
and affected by Blazhko effect with a secondary period of 537 days
(Oosterhoff 1946, Szeidl 1976) or 533 days (Kanyo 1980) and a minor
variation of the secondary period with a periodicity of
about 58 $\div$ 62
days (Kanyo 1980).
We have obtained UBVRI photoelectric light curves and CORAVEL
(Baranne et al. 1979, Mayor 1985)
radial
velocity curves of this star.
{}From our data we have derived
an amplitude in V of 1.16 mag, and confirmed the occurrence
of the Blazhko (Clementini et al. 1995).
Spectra taken at minimum light indicate (Clementini et al. 1991) that
[Fe/H]=$-$0.10 or [Fe/H]=+0.14 if the new metallicity calibration
of the Ca II K line equivalent width by Clementini et al. (1994) is adopted,
placing it indeed among the most metal rich RR Lyraes.
The Baade-Wesselink method has been applied to this star by
Jones et al. (1988) and the derived visual absolute
magnitude is M$_{V}$=0.98 mag.
\bigskip\noindent
4. DISCUSSION AND CONCLUSIONS
\medskip\noindent
The results presented in the previous sections show that the sample
of eight {\it anomalous} {\it ab}-type RR Lyraes selected by CMT on the
basis of data collected from the
1976 version of the GCVS suffer from several uncertainties. From our analysis
we have found that:
\par\noindent
a) In general our derived V$_{max}$ magnitudes are brighter
than those given by the
GCVS$_{t}$ and GCVS4, but this difference can
be simply due to
the fact that these catalogues list the B or the photographic magnitudes
which are generally 0.10 $-$ 0.30 mag fainter than the V
magnitudes at the average colours of RR Lyraes. AT Com is the only
exception being the m$_{max}$ value given from GCVS4 $\sim$ 0.8 mag brighter
that our V$_{max}$. \par\noindent
b) In six cases (AT Com, BE Com, BS Com, CY Com, DL Com and CM Leo) our
amplitude is significantly smaller than that quoted in the GCVS$_{t}$ and
GCVS4. Again, part of the difference can be due to the dependence of the
amplitude on the observed band, the B-amplitude being usually larger
(by 0.15 $-$ 0.35 mag) than the
V-amplitude. This, for instance, could account for the difference of
about 0.25 mag found between the amplitude of the V light curve of DL Com and
the amplitude of the m$_{pg}$ light curve derived by Kurochkin
\& Pochinok (1974), (see Section 3.4).
But in general
the differences are too large and should probably be ascribed on one side to
the extreme difficulty in measuring with the older photographic techniques
the magnitude at minimum light of such faint objects, and on the
other side to the peculiar variations in shape and amplitudes of the
light curve
that most of these variables exhibit.
This problem is particularly relevant when we compare our data with those
by Meinunger \& Wenzel (1968).
\par\noindent
c) In at least three cases (AT Com, CU Com and CM Leo, with the possible
addition of BE Com), the actual pulsation seems to be in the first
harmonic rather than in the fundamental mode. The light curve of
BE Com is fairly noisy and anomalous and we can ascribe to this
reason the misclassification of the
attributed pulsation type, but the other three appear as typical {\it c}-type
RR Lyraes. It must be noted, however, that these stars also show anomalous
variations of the amplitude of their {\it c}-type light curves.
A detailed analysis of the anomalies in the pulsational properties of these
variables and of their causes
is beyond the purposes of the present paper. Here we only list
and discuss briefly some plausible interpretations of these "anomalies".
\par
The scatter and the variable amplitudes of the light curves of CM Leo,
AT, CU, BE and BS Com could be due to the Blazhko effect:
a long term (10-200 days) periodic modulation of amplitude
and/or shape of the light curve, which is
superimposed to the main period variation and has been
detected in roughly 30\% of the known RR Lyraes (Szeidl 1988).
Indeed, Kurochkin (1961) reports BS Com as possibly affected by Blazhko,
and our data seem to confirm this hypothesis suggesting a
Blazhko periodicity of about 27-28 days.
\par
So far, only very few {\it c}-type RR Lyrae stars have been reported
as affected by Blazhko (Szeidl 1976), and the actual occurrence of the
phenomenon for {\it c}-type pulsators has not been definitely confirmed
(see Smith et al. 1994). Szeidl (1988) pointed out that
no RR{\it c} variable shows the
long-term curve modulation of the classic Blazhko effect observed in the
{\it ab}-type RR Lyraes and that probably RR$_{c}$ stars showing
variations in the amplitude of their light curves are rather
double-mode pulsators.
Double-mode RR Lyraes (RR$_{d}$) are variables that pulsate both
in the fundamental
and in the first harmonic mode with a beat period of about one day.
These objects exhibit a very large scatter in their light curves caused
by cycle-to-cycle amplitude changes. CM Leo,
whose amplitude variations seem to occur on a time scale of about 1 day,
could be a good candidate for double-mode pulsation.
\par
Finally, an alternative explanation of the light curve anomalies
might be that some of these stars are
eclipsing binaries
misclassified as RR Lyrae stars.
In his photometric study of 93 poorly studied
variables fainter than the tenth magnitude and selected from the GCVS4,
Schmidt (1991) found that about 7.5\% of the studied variables
are erroneously classified as RR Lyrae stars on the GCVS4.
Schmidt's (1991) re-classification is based on
the appearance of the observed light curves, together with the comparison
of the V and R light curve amplitudes, and the use of the
V {\it vs} V$-$R colour-magnitude
diagram. The latter is suggested by Schmidt to be a very powerful tool
to distinguish the intrinsic pulsational variability from the
geometrical effect of binarity.
We have multicolour photometry only for CU Com, and, as mentioned in
Section 3.2, it suggests that CU Com is an RR Lyrae variable.
For the remaining suspected {\it c}-type stars we only have photometry in the
V band and therefore we cannot check the
location of these stars on the colour-magnitude diagram.

If the variability were produced by eclipses the supposed binary should
be formed by a contact pair (because of the continuous variability of
the light curve and the shortness of the period) and belong to the
W UMa type.
What we know on these systems and the morphology of the corresponding
light curves is enough to make the hypothesis of binarity very
unlikely.
In  Figure 4 we have plotted the data for CM Leo, AT and BE Com
( the variables with a more symmetric light curve) using a period double of
that derived  assuming them to be pulsating stars.
In doing this we have followed Schmidt's (1991) remark that
the "light curves of some eclipsing binary
are difficult to distinguish from Bailey type {\it c} RR Lyraes if a
period half the correct value has been assumed".
The curves have been compared
with a library of contact binary light-curves computed by one of us (CM,
unpublished)
by means of the last released 1994 version of the Wilson and Devinney light
curve synthesis code (in original form in Wilson and Devinney 1971).
All the light curves have too peaked maxima and too flat and slowly
growing minima to be reasonably fitted by any acceptable combination
of parameters (see the {\it best} fit solution obtained for
BE Com shown in panel 4d).
The peaked maxima imply an extreme degree of contact
(contact with the outer lagrangian surface) which is never observed
in real contact binaries. Moreover, in contact binary curves the ratio $r$ of
the phase interval  from the maximum to half curve depth to that
from half depth to the minimum is typically larger than 1, i.e. the decrease
in magnitude is slow moving away from maximum, for it is not due to eclipse
but only to the ``proximity effects" (mainly gravity darkening and
the Roche deformation of the surfaces).
The value of $r$ approaches unity only for deep contact and low value of
the inclination, that however drastically decreases the minimum depth.

To conclude, in the hypothesis of binarity we can only get a low quality
fit of the observed data, (see Fig. 4d), with a set of system parameters rather
improbable for a contact binary configuration. Even if only on a
morphological basis, we are therefore oriented to discard the hypothesis.
\par
The present data are not numerous enough, or sufficiently well
distributed in time to allow us to undoubtedly assess the occurrence and,
in case, to establish the periodicity of the Blazhko effect for our
variables, or to distinguish
between the Blazhko
and double-mode pulsation.
To explore the problem, further additional data in
different colour bands and with appropriate time coverage
are required. In particular, the identification of
RR$_{d}$
variables requires adequate phase coverage of the primary
and secondary period of pulsation, while the identification of Blazhko
objects requires the coverage of the Blazhko period that
could be as long as 10-200 days.
In both cases we therefore need
observations that span several hours on a given night and in a few
subsequent nights, as well as in several months and/or years.
To this
purpose appropriate monitoring of these variables is planned in future
observing runs at the 1.52m telescope.
\bigskip
In any case, with the new data, we can reexamine the issue of possible metal
rich
variables at large distances from the disc raised by CMT.
To this aim we have derived the minimum distance $z$ of each object by
assuming an absolute magnitude of M$_V$=0.6 at maximum light. This is not
an accurate measure of the distance but is sufficient for our purposes.
Note that the ranking of the distances from the plane coincides with that of
the distances from the sun, because all the variables are located in the
direction of the Galactic Pole.
RS Boo is the closest variable with $z$=0.65 and its high
metallicity is consistent with the abundances usually found
at that height. The distances of the other variables range from 2.22 kpc
for BS Com to 9.46 kpc for BE Com. CY Com is the farthest {\it
ab}-type variable
with a distance of 5.37 kpc and its period of 0.758$^d$ suggests that it is
a normal halo metal poor star. DL Com is therefore the only case of
certain {\it ab}-type variable with short period and located as far as 2.66 kpc
and we cannot consider this single object as evidence for a significant
metal rich population at large distances from the galactic plane.
\bigskip\noindent
ACKNOWLEDGEMENTS
\medskip\noindent
We wish to thank Dr. A.Bonifazi and Dr. M.Lolli for discussion and
useful comments on the eclipsing binary interpretation for the anomalous
{\it c}-type light curves.
\bigskip\noindent
REFERENCES
\medskip\noindent
Bloomfield, P. 1976, Fourier Analysis of Time Series, (John Wiley \& Sons eds,
\par New York)
\medskip\noindent
Castellani, V., Maceroni, C., Tosi, M. 1981, A.A. 102, 411
\medskip\noindent
Castellani, V., Maceroni, C., Tosi, M. 1983, A.A. 128, 64, CMT
\medskip\noindent
Clementini, G., Cacciari, C., Prevot, L., Lolli, M. 1995, in preparation
\medskip\noindent
Clementini, G., Carretta, E., Gratton, R., Merighi, R., Mould, J.R.,
  McCarthy, \par J.K. 1994, A.J., submitted
\medskip\noindent
Clementini, G., Tosi, M., Merighi, R. 1991, A.J. 101, 2168
\medskip\noindent
Jones, R.V., Carney, B.W., Latham, D.W. 1988, Ap.J. 332, 206
\medskip\noindent
Kanyo, S. 1980, Inf. Bull. Variable Stars, No. 1832
\medskip\noindent
Kholopov, P.N. et al. 1985, General Catalogue of Variable Stars, 4th ed.
\par (Moscow :
Nauka Publishing House), GCVS4
\medskip\noindent
Kukarkin, B.V. et al. 1976, third supplement to the third edition
of the General \par Catalogue of
Variable Stars, GCVS$_{t}$
\medskip\noindent
Kurochin, N.E. 1961, Perem. Zvezdy 13, N. 5, 331
\medskip\noindent
Kurochin, N.E., Pochinok, B.D., 1974, Perem. Zvezdy, Prilozh. 2, N. 8, 91
\medskip\noindent
Landolt, A.U. 1982, A.J. 88, 439
\medskip\noindent
Landolt, A.U. 1993, A.J. 104, 340
\medskip\noindent
Layden, A.C. 1993, Ph.D. thesis, Yale University, USA
\medskip\noindent
Lub, J. 1977, Ph.D. Thesis, University of Leiden
\medskip\noindent
Mayor, M. 1985, in A.G. Davis Philip, D.W. Latham eds., Stellar Radial
Velocities, \par IAU Coll. 88, p.35
\medskip\noindent
Meinunger, L., Wenzel, W. 1968, VSS 7, H.4, 389
\medskip\noindent
Oosterhoff, P.Th. 1946, Bull. Astr. Inst. Netherlands, 10, 103 (N.369)
\medskip\noindent
Preston, G.W. 1959, Ap.J. 130, 507
\medskip\noindent
Schmidt, E.G., 1991, A.J. 102, 1766.
\medskip\noindent
Schmidt, E.G., Reiswing, D.E. 1993, A.J. 106, 2429.
\medskip\noindent
Smith, H.A. 1990, P.A.S.P. 102, 124
\medskip\noindent
Smith, H.A., Matthews, J.M, Lee, K.M., Williams, J., Silbermann, N.A.,
 Bolte, \par M. 1994, A.J. 107, 679
\medskip\noindent
Spinrad, H. 1959, Ap.J. 130, 539
\medskip\noindent
Szeidl, B. 1976, in W.S. Fitch ed. Multiple Periodic Variable Stars (Reidel,
\par Dordrecht), p. 133
\medskip\noindent
Szeidl, B. 1988, in G.Kovacs, L.Szbados, B.Szeidl eds Multimode Stellar
Pulsations \par (Konkoly Observatory, Budapest), p. 45
\medskip\noindent
Wilson R.E, Devinney E.J 1971, Ap.J. 166, 605
\medskip\noindent
\vfill\eject
\par\noindent
\centerline {FIGURE CAPTIONS}
\medskip\noindent
{\bf Figure 1 :} Finding charts for CM Leo, CU Com, CY Com, DL Com,
 AT Com, BE Com and BS Com, respectively. The variable stars are
 marked by dashes, the comparison stars by the identification names.
Each chart shows a region of the sky of
$10' \times 10'$, north is up and east to the left.
\medskip
\par\noindent
{\bf Figure 2 :} Calibration curve for data taken with the RCA CCD.
\medskip
\par\noindent
{\bf Figure 3 a,b,c,d,e,f,g :} Light curves of CM Leo, CU, CY, DL, AT, BE
and BS Com,
 respectively (upper panels). Bottom panels show the residual around the
average magnitude difference between the two brighter comparison stars of each
variable.
\medskip
\par\noindent
{\bf Figure 4 a,b,c,d :} Light curves of (a) CM Leo, (b) AT Com and (c) BE Com,
 using periods doubled with respect to the values given in Table 4.
Panel 4d shows
the {\it best} fit solution of an eclipsing binary model computed for
BE Com.
\vfill\eject

\newdimen\digitwidth
\setbox0=\hbox{\rm0}
\digitwidth=\wd0
\catcode`?=\active
\def?{\kern\digitwidth}
\vsize=18truecm
\hsize=18truecm
\nopagenumbers

\tabskip=2em plus1em minus1em
\noindent{\bf Table 1.} -- Journal of observations
\trule
\halign to
\hsize{#\hfil&\hfil#\hfil&\hfil#&\hfil#\hfil&\hfil#\hfil\cr
Star\hfil&RA$_{2000}$&DEC$_{2000}$&N. Observations& Observed Interval$^{\rm
a}$\cr
\mrule
CM Leo&11 56 14&21 15 32&?71&9403$-$9494\cr
CU Com&12 24 47&22 24 29&118&7558$-$9472\cr
CY Com&12 28 20&24 57 19&177&7599$-$9471\cr
DL Com&12 34 21&16 08 25&?94&9403$-$9522\cr
AT Com&12 45 44&18 12 11&105&8675$-$9473\cr
BE Com&12 58 02&19 51 34&?68&7652$-$9509\cr
BS Com&13 34 40&24 16 39&?97&9440$-$9522\cr
RS Boo&14 33 33&31 45 14&~ --- &---\cr
\brule }
\medskip\par\noindent
$^{\rm a}$ Observed Intervals are given as JD$-$2,440,000.
\vfill\eject

\newdimen\digitwidth
\setbox0=\hbox{\rm0}
\digitwidth=\wd0
\catcode`?=\active
\def?{\kern\digitwidth}
\vsize=18truecm
\hsize=15truecm
\nopagenumbers

\tabskip=2em plus1em minus1em
\noindent{\bf Table 2.} -- The comparison stars.
\trule
\halign to
\hsize{#\hfil&\hfil#\hfil&\hfil#&\hfil#\hfil&#\hfil\cr
Comparison star&N$_{\rm GSC}$&RA$_{2000}$&DEC$_{2000}$&??V\cr
\mrule
C1 (CM Leo)&1277-1007&11 56 00&21 21 31&12.53\cr
C1 (CU Com)&1664-1247&12 24 49&22 23 14&14.11\cr
C1 (CY Com)&0011-0010&12 28 21&24 55 58&14.79\cr
C3 (DL Com)&1201-1026&12 34 04&16 07 27&14.53\cr
C1 (AT Com)&2144-1839&12 45 48&18 15 00&15.10:\cr
C1 (BE Com)&0159-0159&12 58 05&19 52 52&15.17\cr
C2 (BS Com)&0454-0454&13 34 30&24 18 56&13.91\cr
BD +32 24 87 (RS Boo)&0753-0607&14 32 29&31 43 30&10.65\cr
\brule }
\medskip\par\noindent
- Data for the comparison star of RS Boo are from Clementini et al 1994a.
\vfill\eject

\newdimen\digitwidth
\setbox0=\hbox{\rm0}
\digitwidth=\wd0
\catcode`?=\active
\def?{\kern\digitwidth}
\vsize=18truecm
\hsize=8truecm
\nopagenumbers

\tabskip=2em plus1em minus1em
\noindent{\bf Table 3.} -- The standard stars.
\trule
\halign to
\hsize{#\hfil&\hfil#\hfil&\hfil#\cr
Star&V&B$-$V\cr
\mrule
84971 &?8.636&$-$0.159\cr
101 389&?9.962&0.427\cr
101 324&?9.743&1.157\cr
+5 2468&?9.348&$-$0.116\cr
100340& 10.117&$-$0.242\cr
103 462&10.111&0.564\cr
105 28&?8.345&1.039\cr
105 663&?8.760&0.342\cr
+2 2711&10.367&$-$0.162\cr
107 544&?9.037&0.401\cr
107 347&?9.443&1.296\cr
149382&?8.944&$-$0.281\cr
108 702&?8.208&0.559\cr
108 1491&?9.059&0.965\cr
PG1633+099&14.396&$-$0.192\cr
PG1633+099A&15.254&0.876\cr
PG1633+099B&12.966&1.081\cr
PG1633+099C&13.224&1.133\cr
\brule }
\vfill\eject

\newdimen\digitwidth
\setbox0=\hbox{\rm0}
\digitwidth=\wd0
\catcode`!=\active
\def!{\kern\digitwidth}
\vsize=18truecm
\hsize=19truecm
\hoffset=-1truecm
\nopagenumbers

\tabskip=2em plus1em minus1em
\noindent{\bf Table 4.} -- Derived quantities for the variable stars.
\trule
\halign to
\hsize{#\hfil&\hfil#\hfil&\hfil#\hfil&\hfil#&\hfil#\hfil&\hfil#\hfil&#\hfil&
\hfil#\hfil&
\hfil#\hfil&\hfil#\hfil\cr
Star\hfil&Type&Epoch&Period!!&A$_{V}$&V$_{max}$&Period(GCVS4)&A(GCVS4)&m$_{max}$(GCVS4)\cr
\mrule
CM Leo&c&49403.6122!&0.361479!&0.49&13.47&!!!0.361732&1.10&13.80\cr
CU Com&c&49451.56667&0.405749!&0.58&13.02&!!!0.416091&0.50&13.10\cr
CY Com&ab&47632.52446&0.757880!&0.70&14.25&!!!0.757881&1.10&14.40\cr
DL Com&ab&49437.5242!&0.4320590&1.26&12.72&!!!0.4321025&2.00&12.90\cr
AT Com&c&48675.6303!&0.3444676&0.62&14.72&!!!0.344465&1.60&13.90\cr
BE Com&?&37705.608!!&0.414209!&0.58&15.48&!!!0.41421&0.90&15.70\cr
BS Com&?&49451.60664&0.3629668&0.70&12.33&!!!0.36350&1.20&12.40\cr
RS Boo&ab&46948.7198!&0.3773397&1.16&!9.64&!!!0.37733896&1.15&!9.69\cr
\brule }
\medskip\par\noindent
- Epoch of RS Boo is from Jones {et al.} 1988, Period is from Clementini
{\it et al.} 1994a.
\par\noindent
- Epoch of BE Com is from GCVS$_{t}$, (see Section 3.6).
\vfill\eject

\newdimen\digitwidth
\setbox0=\hbox{\rm0}
\digitwidth=\wd0
\catcode`?=\active
\def?{\kern\digitwidth}
\tenpoint\rm
\singlespace
\vsize=30truecm
\hsize=18truecm
\nopagenumbers
\tabskip=2em plus1em minus1em
\noindent{\bf Table 5a.} V photometry for CM Leo.
\trule
\halign to
\hsize{\hfil#\hfil&\hfil#\hfil&\hfil#\hfil&\hfil#\hfil&\hfil#\hfil&\hfil#
\hfil&\hfil#\hfil&\hfil#\hfil&\hfil#\hfil&\hfil#\hfil&\hfil#\hfil\cr
HJD$-$2440000\hfil&$\phi$&$V$&&HJD$-$2440000\hfil&$
\phi$&$V$&&HJD$-$2440000\hfil&$\phi$&$V$\cr
\mrule
9403.5089&0.714&13.93&&9434.4437&0.293&13.75&&9437.3311&0.281&13.63\cr
9403.5154&0.732&13.87&&9434.4515&0.314&13.81&&9437.3961&0.460&13.92\cr
9403.5224&0.751&13.85&&9434.4596&0.337&13.87&&9437.4042&0.483&13.92\cr
9403.5303&0.774&13.78&&9434.4674&0.358&13.88&&9437.4117&0.503&13.94\cr
9403.5382&0.795&13.70&&9434.4752&0.380&13.94&&9437.4549&0.623&13.93\cr
9403.5453&0.815&13.66&&9434.4833&0.402&13.95&&9437.4624&0.644&13.94\cr
9403.5518&0.833&13.65&&9435.3672&0.848&13.58&&9440.4033&0.779&13.84\cr
9403.5583&0.851&13.64&&9435.3743&0.867&13.55&&9440.4128&0.806&13.75\cr
9403.5651&0.870&13.64&&9435.3813&0.886&13.54&&9441.3160&0.304&13.72\cr
9403.5724&0.890&13.62&&9435.3883&0.906&13.53&&9441.3254&0.330&13.77\cr
9403.5789&0.908&13.62&&9435.3954&0.925&13.52&&9441.3326&0.350&13.81\cr
9403.5854&0.926&13.59&&9435.4024&0.945&13.51&&9441.3410&0.374&13.83\cr
9403.5920&0.944&13.57&&9435.4096&0.965&13.50&&9441.3491&0.396&13.86\cr
9403.5991&0.964&13.58&&9435.4167&0.984&13.49&&9441.3570&0.418&13.88\cr
9403.6059&0.983&13.57&&9435.4236&0.004&13.48&&9441.3643&0.438&13.90\cr
9403.6122&0.000&13.56&&9435.4314&0.025&13.48&&9441.3713&0.457&13.92\cr
9434.3764&0.106&13.58&&9436.3945&0.689&13.91&&9441.3785&0.477&13.94\cr
9434.3852&0.131&13.58&&9436.4039&0.716&13.88&&9442.5285&0.659&13.95\cr
9434.3941&0.156&13.57&&9436.4122&0.738&13.84&&9442.5423&0.697&13.94\cr
9434.4030&0.180&13.61&&9436.4201&0.760&13.80&&9442.5512&0.721&13.93\cr
9434.4111&0.202&13.66&&9436.4279&0.782&13.74&&9494.3780&0.096&13.49\cr
9434.4193&0.225&13.62&&9436.4364&0.805&13.67&&9494.3863&0.119&13.47\cr
9434.4279&0.249&13.70&&9436.4442&0.827&13.60&&9494.3943&0.141&13.47\cr
9434.4359&0.271&13.72&&9437.3203&0.251&13.58\cr
\brule }
\vfill\eject

\newdimen\digitwidth
\setbox0=\hbox{\rm0}
\digitwidth=\wd0
\catcode`?=\active
\def?{\kern\digitwidth}
\tenpoint\rm
\singlespace
\vsize=30truecm
\hsize=18truecm
\nopagenumbers
\tabskip=2em plus1em minus1em
\noindent{\bf Table 5b.} V photometry for CU Com.
\trule
\halign to
\hsize{\hfil#\hfil&\hfil#\hfil&\hfil#\hfil&\hfil#\hfil&\hfil#\hfil&\hfil#
\hfil&\hfil#\hfil&\hfil#\hfil&\hfil#\hfil&\hfil#\hfil&\hfil#\hfil\cr
HJD$-$2440000\hfil&$\phi$&$V$&&HJD$-$2440000\hfil&$
\phi$&$V$&&HJD$-$2440000\hfil&$\phi$&$V$\cr
\mrule
7558.5246&0.450&13.59&&7598.5897&0.194&13.27&&8725.4617&0.458&13.53\cr
7558.5399&0.488&13.61&&7598.5958&0.209&13.29&&8725.4644&0.465&13.54\cr
7558.5520&0.518&13.61&&7598.6024&0.225&13.30&&8725.4663&0.469&13.53\cr
7558.5780&0.582&13.60&&7598.6074&0.238&13.31&&8725.4679&0.473&13.54\cr
7558.5893&0.610&13.61&&7598.6128&0.251&13.33&&8725.4695&0.477&13.53\cr
7558.6149&0.673&13.57&&7598.6227&0.275&13.35&&8725.4710&0.481&13.56\cr
7558.6247&0.697&13.54&&7598.6293&0.292&13.38&&8725.4725&0.484&13.53\cr
7558.6280&0.705&13.53&&7598.6356&0.307&13.39&&8725.4744&0.489&13.54\cr
7558.6313&0.713&13.52&&7598.6402&0.318&13.40&&8725.4945&0.539&13.56\cr
7558.6347&0.722&13.50&&7598.6443&0.329&13.41&&8725.4973&0.545&13.57\cr
7558.6566&0.776&13.41&&7598.6485&0.339&13.42&&8725.4992&0.550&13.58\cr
7558.6614&0.788&13.39&&7598.6531&0.350&13.44&&8725.5007&0.554&13.57\cr
7558.6654&0.798&13.38&&7598.6573&0.361&13.44&&8725.5023&0.558&13.58\cr
7558.6691&0.807&13.37&&7598.6615&0.371&13.46&&8725.5042&0.563&13.58\cr
7558.6727&0.815&13.35&&7598.6662&0.382&13.46&&8725.5059&0.567&13.57\cr
7558.6854&0.847&13.33&&7598.6703&0.393&13.46&&8725.5080&0.572&13.58\cr
7558.6892&0.856&13.30&&7598.6745&0.403&13.48&&8725.5104&0.578&13.60\cr
7558.6927&0.865&13.28&&7598.6787&0.413&13.49&&8725.5119&0.582&13.59\cr
7598.4233&0.784&13.46&&7598.6837&0.426&13.50&&8725.5134&0.585&13.57\cr
7598.4287&0.797&13.43&&8696.5628&0.234&13.34&&8725.5142&0.587&13.59\cr
7598.4388&0.822&13.37&&8696.5672&0.245&13.35&&9076.4956&0.608&13.58\cr
7598.4485&0.846&13.32&&8696.5709&0.254&13.34&&9076.4972&0.612&13.59\cr
7598.4527&0.856&13.29&&8696.5743&0.262&13.36&&9076.4987&0.616&13.60\cr
7598.4571&0.867&13.28&&8696.5769&0.269&13.36&&9076.5000&0.619&13.59\cr
7598.4614&0.878&13.27&&8696.5796&0.275&13.35&&9076.5015&0.623&13.60\cr
7598.4659&0.889&13.25&&8696.5837&0.286&13.36&&9076.5029&0.626&13.60\cr
7598.4705&0.900&13.23&&8696.5865&0.293&13.37&&9451.3272&0.410&13.42\cr
7598.4753&0.912&13.23&&8696.5903&0.302&13.38&&9451.3360&0.431&13.43\cr
7598.4807&0.925&13.23&&8696.5987&0.323&13.38&&9451.4489&0.710&13.53\cr
7598.4872&0.941&13.21&&8696.6019&0.331&13.39&&9451.4571&0.730&13.51\cr
7598.5000&0.973&13.16&&8696.6045&0.337&13.41&&9451.4981&0.831&13.27\cr
7598.5066&0.989&13.15&&8696.6072&0.344&13.42&&9451.5069&0.853&13.18\cr
7598.5412&0.074&13.19&&8696.6100&0.350&13.41&&9451.5580&0.979&13.03\cr
7598.5486&0.093&13.19&&8696.6146&0.362&13.43&&9451.5667&0.000&13.02\cr
7598.5538&0.106&13.20&&8696.6174&0.369&13.43&&9451.5755&0.022&13.03\cr
7598.5583&0.117&13.20&&8725.4526&0.435&13.51&&9451.5843&0.043&13.04\cr
7598.5628&0.128&13.21&&8725.4550&0.441&13.50&&9472.5555&0.729&13.59\cr
7598.5695&0.144&13.23&&8725.4565&0.445&13.51&&9472.5656&0.753&13.58\cr
7598.5748&0.157&13.24&&8725.4579&0.448&13.51\cr
7598.5842&0.181&13.26&&8725.4598&0.453&13.51\cr
\brule }
\vfill\eject

\newdimen\digitwidth
\setbox0=\hbox{\rm0}
\digitwidth=\wd0
\catcode`?=\active
\def?{\kern\digitwidth}
\tenpoint\rm
\singlespace
\voffset=-2.5truecm
\vsize=30truecm
\hsize=18truecm
\nopagenumbers
\tabskip=2em plus1em minus1em
\noindent{\bf Table 5c.} V photometry for CY Com.
\trule
\halign to
\hsize{\hfil#\hfil&\hfil#\hfil&\hfil#\hfil&\hfil#\hfil&\hfil#\hfil&\hfil#\hfil&\hfil#\hfil&
\hfil#\hfil&\hfil#\hfil&\hfil#\hfil&\hfil#\hfil\cr
HJD$-$2440000\hfil&$\phi$&$V$&&HJD$-$2440000\hfil&$\phi$&$V$&&HJD$-$2440000\hfil&$\phi$&$V$\cr
\mrule
7599.3526&0.231&14.53&&7655.5339&0.360&14.66&&8674.5127&0.872&14.71\cr
7599.3608&0.242&14.55&&7655.5391&0.367&14.66&&8674.5182&0.880&14.65\cr
7599.3686&0.252&14.56&&7655.5462&0.376&14.68&&8674.5224&0.885&14.58\cr
7599.3763&0.262&14.57&&7655.5540&0.387&14.66&&8674.5266&0.891&14.52\cr
7599.3839&0.272&14.56&&7655.5632&0.399&14.69&&8674.5307&0.896&14.50\cr
7599.3915&0.282&14.59&&7655.5686&0.406&14.71&&8674.5349&0.902&14.47\cr
7599.3992&0.292&14.60&&7655.5831&0.425&14.73&&8674.5408&0.909&14.45\cr
7599.4072&0.303&14.61&&7655.5885&0.432&14.73&&8674.5429&0.912&14.47\cr
7599.4147&0.313&14.63&&7655.5945&0.440&14.73&&8674.5478&0.919&14.45\cr
7599.4369&0.342&14.67&&7655.6000&0.448&14.73&&8674.5512&0.923&14.44\cr
7599.4478&0.356&14.69&&7948.4220&0.817&14.94&&8674.5554&0.929&14.41\cr
7599.4554&0.366&14.67&&7948.4313&0.830&14.92&&8674.5582&0.932&14.40\cr
7599.4630&0.376&14.68&&7948.4395&0.840&14.89&&8674.5623&0.938&14.38\cr
7599.4705&0.386&14.70&&7948.4479&0.851&14.85&&8674.5658&0.942&14.36\cr
7599.4784&0.397&14.71&&7948.4566&0.863&14.79&&8675.4468&0.105&14.39\cr
7599.4859&0.407&14.75&&7948.4646&0.874&14.72&&8675.4517&0.111&14.39\cr
7599.4935&0.417&14.75&&7948.4730&0.885&14.61&&8675.4565&0.118&14.41\cr
7599.5011&0.427&14.74&&7948.5082&0.931&14.42&&8675.4614&0.124&14.41\cr
7632.3717&0.798&14.93&&7948.5165&0.942&14.38&&8675.4663&0.131&14.42\cr
7632.3836&0.814&14.93&&7948.5244&0.953&14.32&&8679.4186&0.346&14.65\cr
7632.3931&0.827&14.92&&7948.5324&0.963&14.29&&8679.4247&0.354&14.65\cr
7632.4009&0.837&14.86&&7948.5404&0.974&14.23&&8679.4316&0.363&14.65\cr
7632.4091&0.848&14.83&&7948.5488&0.985&14.27&&8679.4370&0.370&14.68\cr
7632.4175&0.859&14.80&&7948.5568&0.995&14.25&&8679.4430&0.378&14.68\cr
7632.4253&0.869&14.74&&7948.5653&0.006&14.26&&8679.4478&0.384&14.69\cr
7632.4330&0.879&14.65&&7948.5740&0.018&14.27&&8696.3854&0.733&14.95\cr
7632.4417&0.891&14.54&&7948.5820&0.028&14.28&&8696.3921&0.742&14.94\cr
7632.4497&0.901&14.48&&7993.3568&0.107&14.37&&8696.3973&0.748&14.94\cr
7632.4573&0.911&14.47&&7994.4026&0.487&14.76&&8696.4065&0.761&14.96\cr
7632.4942&0.960&14.31&&7994.4121&0.500&14.77&&8696.4107&0.766&14.96\cr
7632.5021&0.971&14.29&&7994.4218&0.513&14.77&&8696.4148&0.772&14.95\cr
7632.5089&0.979&14.27&&7994.4297&0.523&14.77&&8696.6345&0.061&14.34\cr
7632.5164&0.989&14.24&&7994.4376&0.534&14.78&&8696.6421&0.071&14.35\cr
7632.5245&0.000&14.25&&7994.4454&0.544&14.77&&8696.6464&0.077&14.36\cr
7632.5326&0.011&14.25&&7994.4536&0.555&14.79&&8696.6506&0.083&14.36\cr
7632.5420&0.023&14.25&&7994.4618&0.565&14.79&&8696.6560&0.090&14.37\cr
7632.5504&0.034&14.28&&7994.4706&0.577&14.80&&8696.6609&0.096&14.39\cr
7632.5584&0.045&14.29&&7994.4796&0.589&14.80&&8696.6652&0.102&14.39\cr
7632.5664&0.055&14.32&&7994.4883&0.600&14.82&&8696.6689&0.107&14.40\cr
7632.5743&0.066&14.31&&7994.4976&0.613&14.81&&8696.6728&0.112&14.40\cr
7655.3642&0.136&14.44&&7994.5063&0.624&14.82&&8696.6770&0.118&14.41\cr
7655.3738&0.149&14.46&&7994.5150&0.636&14.82&&8725.5315&0.190&14.49\cr
7655.3829&0.161&14.47&&7994.5230&0.646&14.84&&8725.5368&0.197&14.50\cr
7655.3910&0.172&14.48&&7994.5314&0.657&14.84&&8725.5409&0.202&14.50\cr
7655.3975&0.180&14.48&&7994.5392&0.668&14.86&&8725.5451&0.208&14.50\cr
7655.4035&0.188&14.49&&7994.5470&0.678&14.87&&8725.5491&0.213&14.51\cr
7655.4109&0.198&14.49&&7994.5565&0.690&14.89&&8762.3872&0.820&14.94\cr
7655.4208&0.211&14.52&&7994.5643&0.701&14.89&&8762.3918&0.826&14.93\cr
7655.4267&0.219&14.53&&7994.5732&0.712&14.91&&8762.3960&0.832&14.91\cr
7655.4630&0.267&14.55&&7994.5818&0.724&14.90&&8762.4005&0.838&14.90\cr
7655.4709&0.277&14.57&&7995.3636&0.755&14.94&&9436.4613&0.241&14.50\cr
7655.4817&0.291&14.57&&7995.3719&0.766&14.95&&9436.4707&0.253&14.51\cr
7655.4891&0.301&14.59&&7995.3796&0.777&14.96&&9436.4806&0.266&14.52\cr
7655.4961&0.310&14.60&&8674.4787&0.827&14.92&&9436.5534&0.362&14.63\cr
7655.5009&0.317&14.61&&8674.4856&0.837&14.88&&9436.5669&0.380&14.64\cr
7655.5060&0.323&14.62&&8674.4891&0.841&14.86&&9436.5804&0.398&14.67\cr
7655.5119&0.331&14.62&&8674.4988&0.854&14.82&&9436.6460&0.484&14.74\cr
7655.5177&0.339&14.62&&8674.5030&0.859&14.78&&9442.4602&0.156&14.42\cr
7655.5243&0.348&14.64&&8674.5078&0.866&14.76&&9471.4730&0.438&14.69\cr
\brule }
\vfill\eject

\newdimen\digitwidth
\setbox0=\hbox{\rm0}
\digitwidth=\wd0
\catcode`?=\active
\def?{\kern\digitwidth}
\tenpoint\rm
\singlespace
\vsize=30truecm
\hsize=18truecm
\nopagenumbers
\tabskip=2em plus1em minus1em
\noindent{\bf Table 5d.} V photometry for DL Com.
\trule
\halign to
\hsize{\hfil#\hfil&\hfil#\hfil&\hfil#\hfil&\hfil#\hfil&\hfil#\hfil&\hfil#
\hfil&\hfil#\hfil&\hfil#\hfil&\hfil#\hfil&\hfil#\hfil&\hfil#\hfil\cr
HJD$-$2440000\hfil&$\phi$&$V$&&HJD$-$2440000\hfil&$
\phi$&$V$&&HJD$-$2440000\hfil&$\phi$&$V$\cr
\mrule
9403.6371&0.568&13.85&&9435.4480&0.195&13.32&&9440.5619&0.031&12.80\cr
9403.6517&0.602&13.83&&9435.4550&0.211&13.36&&9441.4799&0.156&13.22\cr
9403.6591&0.619&13.84&&9435.4620&0.227&13.41&&9441.4875&0.173&13.26\cr
9403.6657&0.635&13.85&&9435.4723&0.251&13.44&&9441.5155&0.238&13.40\cr
9403.6734&0.652&13.84&&9435.4795&0.268&13.47&&9441.5238&0.257&13.44\cr
9403.6813&0.671&13.85&&9435.4865&0.284&13.51&&9441.5307&0.273&13.47\cr
9403.6885&0.687&13.86&&9435.4936&0.300&13.54&&9441.5382&0.290&13.50\cr
9403.6960&0.705&13.86&&9435.5007&0.317&13.55&&9441.5498&0.317&13.55\cr
9403.7034&0.722&13.85&&9435.5077&0.333&13.58&&9441.5586&0.337&13.60\cr
9403.7107&0.739&13.85&&9435.5151&0.350&13.61&&9441.5683&0.360&13.61\cr
9406.4884&0.168&13.27&&9435.5225&0.367&13.63&&9441.5752&0.376&13.65\cr
9406.4975&0.189&13.32&&9435.5296&0.384&13.68&&9441.5868&0.403&13.69\cr
9406.5045&0.205&13.37&&9435.5372&0.401&13.70&&9441.5945&0.421&13.73\cr
9406.5128&0.224&13.40&&9435.5442&0.417&13.72&&9441.6022&0.438&13.75\cr
9406.5209&0.243&13.42&&9435.5512&0.434&13.75&&9441.6119&0.461&13.79\cr
9406.5315&0.267&13.48&&9435.5583&0.450&13.76&&9442.4148&0.319&13.54\cr
9434.5087&0.021&12.80&&9435.5654&0.466&13.78&&9442.4372&0.371&13.64\cr
9434.5178&0.042&12.82&&9435.5855&0.513&13.81&&9442.5146&0.550&13.79\cr
9434.5256&0.060&12.87&&9435.5924&0.529&13.81&&9465.5333&0.827&13.95\cr
9434.5339&0.079&12.95&&9436.5989&0.858&13.98&&9465.5415&0.846&13.96\cr
9434.5417&0.097&13.00&&9436.6080&0.880&13.95&&9465.5502&0.866&13.97\cr
9434.5510&0.119&13.06&&9436.6162&0.899&13.85&&9465.5584&0.885&13.93\cr
9434.5590&0.137&13.10&&9436.6241&0.917&13.63&&9473.4336&0.112&13.08\cr
9434.5668&0.155&13.18&&9436.6319&0.935&13.28&&9473.4427&0.133&13.15\cr
9434.5746&0.173&13.21&&9437.4870&0.914&13.76&&9473.4779&0.215&13.35\cr
9434.5823&0.191&13.25&&9437.4951&0.933&13.36&&9473.4863&0.234&13.38\cr
9434.5902&0.209&13.30&&9437.5022&0.949&13.15&&9510.3668&0.594&13.79\cr
9434.5980&0.227&13.35&&9437.5098&0.967&12.84&&9520.3565&0.715&13.89\cr
9434.6061&0.246&13.40&&9437.5290&0.011&12.74&&9522.3896&0.421&13.72\cr
9434.6140&0.264&13.43&&9440.4258&0.716&13.83&&9522.3982&0.441&13.74\cr
9434.6218&0.282&13.45&&9440.4344&0.736&13.82\cr
9434.6297&0.301&13.49&&9440.5528&0.010&12.72\cr
\brule }
\vfill\eject

\newdimen\digitwidth
\setbox0=\hbox{\rm0}
\digitwidth=\wd0
\catcode`?=\active
\def?{\kern\digitwidth}
\tenpoint\rm
\singlespace
\vsize=30truecm
\hsize=18truecm
\nopagenumbers
\tabskip=2em plus1em minus1em
\noindent{\bf Table 5e.} V photometry for AT Com.
\trule
\halign to
\hsize{\hfil#\hfil&\hfil#\hfil&\hfil#\hfil&\hfil#\hfil&\hfil#\hfil&\hfil#\hfil&\hfil#\hfil&
\hfil#\hfil&\hfil#\hfil&\hfil#\hfil&\hfil#\hfil\cr
HJD$-$2440000\hfil&$\phi$&$V$&&HJD$-$2440000\hfil&$\phi$&$V$&&HJD$-$2440000\hfil&$\phi$&$V$\cr
\mrule
8675.4911&0.596&15.30&&8675.6765&0.134&14.85&&8725.3204&0.252&15.01\cr
8675.4977&0.615&15.34&&8675.6799&0.144&14.86&&8725.3244&0.264&14.98\cr
8675.5032&0.631&15.34&&8675.6852&0.159&14.88&&8725.3285&0.275&15.03\cr
8675.5077&0.644&15.34&&8675.6890&0.170&14.90&&8725.3625&0.374&15.14\cr
8675.5136&0.661&15.34&&8675.6935&0.183&14.91&&8725.3672&0.388&15.16\cr
8675.5178&0.673&15.32&&8675.6973&0.195&14.94&&8725.3714&0.400&15.16\cr
8675.5234&0.690&15.32&&8675.7015&0.207&14.92&&8725.3754&0.412&15.21\cr
8675.5275&0.702&15.34&&8696.4350&0.397&15.15&&8725.3808&0.427&15.22\cr
8675.5331&0.718&15.30&&8696.4431&0.420&15.18&&8725.3859&0.442&15.24\cr
8675.5383&0.733&15.28&&8696.4485&0.436&15.20&&8725.3914&0.458&15.25\cr
8675.5466&0.757&15.28&&8696.4545&0.453&15.22&&8725.3967&0.473&15.28\cr
8675.5511&0.770&15.28&&8696.4594&0.468&15.21&&8725.4016&0.488&15.26\cr
8675.5560&0.784&15.21&&8696.4653&0.485&15.26&&8725.4069&0.503&15.31\cr
8675.5629&0.804&15.17&&8696.4693&0.496&15.27&&8725.4116&0.517&15.31\cr
8675.5699&0.825&15.08&&8696.4754&0.514&15.29&&8725.5753&0.992&14.73\cr
8675.5754&0.841&15.01&&8696.4803&0.528&15.33&&8725.5810&0.008&14.72\cr
8675.5806&0.856&14.94&&8696.4862&0.545&15.31&&8725.5860&0.023&14.75\cr
8675.5848&0.868&14.89&&8696.4915&0.561&15.30&&8725.5907&0.037&14.77\cr
8675.5897&0.882&14.87&&8696.4963&0.575&15.33&&8725.5954&0.050&14.78\cr
8675.5938&0.894&14.84&&8696.5014&0.589&15.31&&8725.6001&0.064&14.78\cr
8675.5983&0.907&14.80&&8696.5086&0.610&15.30&&8725.6048&0.077&14.78\cr
8675.6056&0.928&14.78&&8696.5150&0.629&15.29&&8725.6095&0.091&14.80\cr
8675.6105&0.943&14.78&&8696.5211&0.647&15.29&&9076.5765&0.959&14.74\cr
8675.6157&0.958&14.76&&8696.5265&0.662&15.28&&9076.5792&0.967&14.72\cr
8675.6240&0.982&14.75&&8696.5321&0.679&15.27&&9076.5820&0.975&14.73\cr
8675.6303&1.000&14.72&&8696.5371&0.693&15.29&&9076.5849&0.983&14.71\cr
8675.6345&0.012&14.73&&8696.5419&0.707&15.27&&9076.5872&0.990&14.72\cr
8675.6397&0.027&14.74&&8696.5467&0.721&15.27&&9076.5897&0.998&14.71\cr
8675.6438&0.039&14.74&&8704.4717&0.728&15.28&&9436.5115&0.861&14.92\cr
8675.6484&0.052&14.76&&8704.4768&0.742&15.26&&9436.5255&0.902&14.81\cr
8675.6525&0.065&14.77&&8704.4814&0.755&15.19&&9436.5369&0.935&14.76\cr
8675.6574&0.079&14.79&&8704.4860&0.769&15.20&&9436.6659&0.310&15.02\cr
8675.6619&0.092&14.79&&8704.4906&0.782&15.21&&9473.5227&0.306&14.99\cr
8675.6661&0.104&14.82&&8725.3100&0.222&14.95&&9473.5339&0.339&15.02\cr
8675.6716&0.120&14.84&&8725.3163&0.240&14.98&&9473.5450&0.371&15.07\cr
\brule }
\vfill\eject

\newdimen\digitwidth
\setbox0=\hbox{\rm0}
\digitwidth=\wd0
\catcode`?=\active
\def?{\kern\digitwidth}
\tenpoint\rm
\singlespace
\vsize=30truecm
\hsize=18truecm
\nopagenumbers
\tabskip=2em plus1em minus1em
\noindent{\bf Table 5f.} V photometry for BE Com.
\trule
\halign to
\hsize{\hfil#\hfil&\hfil#\hfil&\hfil#\hfil&\hfil#\hfil&\hfil#\hfil&\hfil#
\hfil&\hfil#\hfil&\hfil#\hfil&\hfil#\hfil&\hfil#\hfil&\hfil#\hfil\cr
HJD$-$2440000\hfil&$\phi$&$V$&&HJD$-$2440000\hfil&$
\phi$&$V$&&HJD$-$2440000\hfil&$\phi$&$V$\cr
\mrule
7652.4046&0.956&15.53&&9076.4724&0.997&15.48&&9465.5060&0.218&15.79\cr
7652.4148&0.980&15.55&&9076.4767&0.008&15.48&&9471.4736&0.625&16.01\cr
7652.4303&0.018&15.50&&9076.4806&0.017&15.48&&9473.3896&0.251&15.73\cr
7652.5577&0.325&15.96&&9076.4850&0.028&15.47&&9473.4005&0.277&15.78\cr
7652.5808&0.381&16.03&&9076.5253&0.125&15.61&&9473.4103&0.301&15.82\cr
7653.4723&0.533&16.06&&9076.5289&0.134&15.61&&9473.4203&0.325&15.89\cr
7653.4914&0.579&16.08&&9076.5327&0.143&15.63&&9473.4548&0.408&16.01\cr
7654.3573&0.670&16.00&&9076.5366&0.152&15.65&&9473.4649&0.433&16.03\cr
7654.3704&0.702&16.01&&9076.5404&0.162&15.66&&9473.5079&0.536&15.98\cr
7654.3864&0.740&15.98&&9076.5452&0.173&15.67&&9487.4772&0.262&15.94\cr
7654.4044&0.784&15.92&&9076.5491&0.183&15.70&&9508.4373&0.865&15.54\cr
7654.4665&0.934&15.52&&9076.6007&0.307&15.93&&9508.4471&0.888&15.56\cr
7654.4964&0.006&15.55&&9076.6047&0.317&15.95&&9508.4568&0.911&15.58\cr
7654.5190&0.060&15.63&&9076.6125&0.336&16.00&&9508.4679&0.938&15.60\cr
9076.4233&0.879&15.69&&9076.6171&0.347&16.00&&9508.4783&0.963&15.64\cr
9076.4260&0.885&15.66&&9076.6208&0.356&16.00&&9509.4259&0.251&15.85\cr
9076.4290&0.893&15.61&&9076.6247&0.365&16.00&&9509.4378&0.280&15.94\cr
9076.4324&0.901&15.58&&9076.6285&0.374&16.02&&9509.4478&0.304&15.97\cr
9076.4355&0.908&15.52&&9437.3612&0.270&15.69&&9509.4577&0.328&16.02\cr
9076.4389&0.917&15.51&&9437.5440&0.711&15.96&&9509.4674&0.351&15.99\cr
9076.4421&0.924&15.50&&9437.5622&0.755&16.06&&9509.4782&0.377&16.00\cr
9076.4622&0.973&15.51&&9465.4291&0.032&15.50&&9509.4868&0.398&16.06\cr
9076.4678&0.986&15.49&&9465.4711&0.134&15.66\cr
\brule }
\vfill\eject

\newdimen\digitwidth
\setbox0=\hbox{\rm0}
\digitwidth=\wd0
\catcode`?=\active
\def?{\kern\digitwidth}
\tenpoint\rm
\singlespace
\vsize=30truecm
\hsize=18truecm
\nopagenumbers
\tabskip=2em plus1em minus1em
\noindent{\bf Table 5g.} V photometry for BS Com.
\trule
\halign to
\hsize{\hfil#\hfil&\hfil#\hfil&\hfil#\hfil&\hfil#\hfil&\hfil#\hfil&\hfil#
\hfil&\hfil#\hfil&\hfil#\hfil&\hfil#\hfil&\hfil#\hfil&\hfil#\hfil\cr
HJD$-$2440000\hfil&$\phi$&$V$&&HJD$-$2440000\hfil&$
\phi$&$V$&&HJD$-$2440000\hfil&$\phi$&$V$\cr
\mrule
9440.5769&0.612&13.01&&9465.4527&0.147&12.66&&9492.4338&0.482&12.96\cr
9440.5875&0.641&13.01&&9465.4592&0.165&12.67&&9492.4440&0.510&12.99\cr
9440.5979&0.670&13.01&&9465.4878&0.244&12.73&&9492.4569&0.545&12.99\cr
9440.6091&0.701&13.02&&9465.4949&0.263&12.74&&9492.4648&0.567&13.00\cr
9440.6193&0.729&13.00&&9465.5686&0.466&12.94&&9492.4770&0.601&13.02\cr
9440.6630&0.849&13.01&&9465.5765&0.488&12.95&&9492.4909&0.639&13.02\cr
9441.6424&0.548&13.03&&9465.5816&0.502&12.98&&9492.5018&0.669&13.04\cr
9441.6519&0.574&13.03&&9465.5889&0.522&13.00&&9492.5121&0.697&13.04\cr
9441.6613&0.600&13.05&&9465.5988&0.549&13.01&&9492.5225&0.726&13.03\cr
9442.6545&0.336&12.87&&9465.6059&0.569&13.02&&9492.5341&0.758&13.00\cr
9451.3545&0.305&12.83&&9465.6139&0.591&13.02&&9492.5448&0.788&12.98\cr
9451.3633&0.330&12.87&&9465.6224&0.614&13.03&&7994.5992&0.838&12.97\cr
9451.4265&0.504&13.05&&9465.6302&0.636&13.03&&7994.6223&0.902&12.67\cr
9451.4374&0.534&13.06&&9465.6383&0.658&13.04&&7994.6270&0.915&12.62\cr
9451.4729&0.631&13.10&&9471.3461&0.384&12.81&&7994.6331&0.932&12.58\cr
9451.4824&0.658&13.11&&9471.3555&0.409&12.82&&7994.6392&0.949&12.55\cr
9451.5344&0.801&12.99&&9471.3648&0.435&12.85&&7994.6419&0.956&12.52\cr
9451.5460&0.833&12.90&&9471.3762&0.467&12.90&&7994.6472&0.971&12.51\cr
9451.5987&0.978&12.34&&9471.3848&0.490&12.92&&7994.6499&0.978&12.49\cr
9451.6066&1.000&12.34&&9471.4092&0.557&13.00&&9508.4051&0.484&12.91\cr
9451.6136&0.019&12.35&&9471.4178&0.581&13.02&&9508.4176&0.518&12.95\cr
9451.6198&0.036&12.34&&9471.4511&0.673&13.03&&9511.3458&0.586&12.99\cr
9451.6267&0.055&12.36&&9471.4906&0.782&13.02&&9511.3636&0.635&13.07\cr
9451.6329&0.072&12.38&&9471.4988&0.804&13.03&&9511.4340&0.829&13.06\cr
9451.6394&0.090&12.41&&9471.5379&0.912&12.70&&9511.4511&0.876&13.04\cr
9451.6461&0.109&12.43&&9471.5501&0.946&12.55&&9511.4578&0.894&13.02\cr
9465.3622&0.898&12.52&&9487.4908&0.864&12.99&&9519.3776&0.714&12.99\cr
9465.3741&0.930&12.49&&9492.3569&0.270&12.76&&9519.4472&0.906&12.44\cr
9465.3871&0.966&12.48&&9492.3710&0.309&12.79&&9519.4554&0.928&12.43\cr
9465.3947&0.987&12.50&&9492.3820&0.339&12.81&&9522.4136&0.078&12.39\cr
9465.4033&0.011&12.52&&9492.3920&0.367&12.83&&9522.4228&0.104&12.42\cr
9465.4103&0.030&12.53&&9492.4066&0.407&12.86\cr
9465.4458&0.128&12.63&&9492.4224&0.450&12.92\cr
\brule }
\vfill\eject
\bye